# High-resolution and reliable automatic target recognition based on photonic ISAR imaging system with explainable deep learning

Xiuting Zou[1], Anyi Deng[1], Yiheng Hu[1], Shiyu Hua[1], Linbo Zhang[1], Shaofu Xu[1], and Weiwen Zou[1*]

[1]State Key Laboratory of Advanced Optical Communication Systems and Networks, Intelligent Microwave Lightwave Integration Innovation Center (imLic), Department of Electronic Engineering. Shanghai Jiao Tong University, Shanghai 200240, China.

e-mail: *wzou@sjtu.edu.cn

**Abstract**

Automatic target recognition (ATR) based on inverse synthetic aperture radar (ISAR) images, which is extensively utilized to surveil environment in military and civil fields, must be high-precision and reliable. Photonic technologies' advantage of broad bandwidth enables ISAR systems to realize high-resolution imaging, which is in favor of achieving high-performance ATR. Deep learning (DL) algorithms have achieved excellent recognition accuracies. However, the lack of interpretability of DL algorithms causes the head-scratching problem of credibility. In this paper, we exploit the inner relationship between a photonic ISAR imaging system and behaviors of a convolutional neural network (CNN) to deeply comprehend the intelligent recognition. Specifically, we manipulate imaging physical process and analyze network outputs, the relevance between the ISAR image and network output, and the visualization of features in the network output layer. Consequently, the broader imaging bandwidths and appropriate imaging angles lead to more detailed structural and contour features and the bigger discrepancy among ISAR images of different targets, which contributes to the CNN recognizing and distinguishing objects according to physical laws. Then, based on the photonic ISAR imaging system and the explainable CNN, we accomplish a high-accuracy and reliable ATR. To the best of our knowledge, there is no precedent of explaining the DL algorithms by exploring the influence of the physical process of data generation on network behaviors. It is anticipated that this work can not only inspire the accomplishment of a high-performance ATR but also bring new insights to explore network behaviors and thus achieve better intelligent abilities.

**Introduction**

To achieve quick and correct situation assessment for environment surveillance and monitoring in military and civil fields[1-6], automatic target recognition (ATR) with high performance of accuracy and reliability is urgently required. Inverse synthetic aperture radar (ISAR) has been universally adopted as an efficient and powerful imaging technology for ATR, due to radar's immense advantages of imaging moving targets with all-day, all-weather, and long-range detection capability. Traditionally, due to the limitation of bandwidth[7,8], electronic ISAR systems have a difficulty in obtaining high-resolution enough images which are usually essential to characterize target features, largely preventing the achievement of high-performance ATR. Introducing photonic technologies with the merit of ultrawide bandwidth[7-9] into ISAR systems can overcome the bottleneck of bandwidth and achieve high-resolution imaging, which has been widely studied[10,11]. In the past decade, deep learning (DL)[12] has crossed with a variety of disciplines such as mathematics[13], biomedicine[14], and optoelectronics[15-19], making substantial advances in each other. Thanks to DL algorithms' excellent ability of feature extraction and analysis, DL-powered ATRs[11,20,21], besides the ATR based on ISAR images, have achieved high-precision recognition to which the conventional methods[22,23] far reach. Whereas, the "black-box"[24-26] nature that DL algorithms lack transparency and the tendency of "shortcut learning"[27] which DL algorithms tend to converge according to "shortcut" strategies instead of physical laws, together call out problems that how the DL algorithm makes the decision and whether the reasoning process is reliable.

Recently more and more attention has been drawn into the understanding and explanation of DL algorithms. Abundant researches and works of interpretable DL spring up. In general, those methods can be classified as three categories: the intrinsic explanation[28-31], the post-hoc explanation[32-38], and the theory explanation[39-41]. For the ATR, there are several works[28,29,42] of the intrinsic explanation which introduce a regularization loss to higher network layers to learn semantic representations. Most studies[43-45] focus on the post-hoc explanation to give the relevance between each input features and the decision result of neural networks (NN). Nevertheless, all these methods interpret the NN from the view of data and NN itself, almost neglecting the physical process of data generation. In contrast, exploring the relationship between the hardware systems or devices which generate data and the behaviors of the NNs may be a potentially powerful angle to deeply comprehend the NN.

Here, we accomplish a high-accuracy and credible ATR based on a photonic ISAR imaging system with the explainable NN through exploiting how the physical process of imaging influences on the intelligent recognition behaviors. Concretely, a photonic ISAR imaging system generates ISAR images for three targets. A convolutional neural network is adopted to automatically recognize targets after training. We adjust the imaging bandwidths of the photonic ISAR imaging system and change the imaging angles of targets to produce several different ISAR images of targets. These ISAR images are dropped into the CNN in training or testing phase. We use the layer-wise relevance propagation (LRP)[35], a post-hoc interpretability technology, to obtain the relevance between each pixel of these ISAR images and the corresponding network outputs. In addition, we adopt the t-distribution stochastic neighbor embedding (t-SNE)[46], a dimensionality reduction algorithm, to visualize the features in the network output layer of ISAR images obtained under different imaging physical process. Through analyzing the network recognition results, the results of LRPs, and the visualization of the t-SNE, we draw a conclusion that when the broader imaging bandwidth brings more high-resolution ISAR images and when the appropriate imaging angles keep target's peculiar features in ISAR images, the CNN depends on the major structural and contour features of targets to recognize different objects. Moreover, the broader imaging bandwidths and appropriate imaging angles introduce the bigger discrepancy among different targets' ISAR images. Hence, imaging under broader bandwidths and appropriate angles makes the NN easier to converge and distinguish ISAR images according to targets' peculiar characteristics. Therefore, we have achieved a high-accuracy and reliable ATR based on the photonic ISAR system with the explainable CNN. To the best of our knowledge, it is the first time explaining NN for ATR from the view of exploiting the impact of physical process of data generation on network behaviors according to the domain knowledges of imaging and thus improving the performance of network output. The insights provided by this work would inspire investigations on high-performance ATR and explainable artificial intelligence (AI).

The general ideal of exploring and explaining the intelligent behaviors of the CNN for ATR from the perspective of the physical process of ISAR imaging is depicted in Fig. 1a. There are three parts: a photonic ISAR imaging system, a CNN for ATR, and a post-hoc interpretability technology and the t-SNE. The first one is used to generate abundant data, i.e., ISAR images, whose simplified structure is shown in Fig. 1b. The second one after training automatically recognizes the targets' ISAR images. In the third one, the post-hoc interpretability technology is adopted to obtain the relevance between each pixel of ISAR images and recognition results and the t-SNE is utilized to visualize the features in the network output layer. Figure 1c presents the illustration of the post-hoc interpretability technologies. We change the bandwidth of the photonic ISAR imaging system and the imaging angles of targets to obtain abundant ISAR images. A portion of ISAR images with different imaging resolutions are respectively used to train several CNNs with the same structure and hyperparameters, obtaining well-trained CNNs. By analyzing the recognition outcomes of the CNN, the results of the LRP, and the visualization of the t-SNE, we explore the influence of imaging bandwidths and imaging angles on the behaviors of the CNN during the process of target recognition. This brings deep insights to the intelligent behaviors of the CNN for ATR. The above analysis results of the ISAR images collected under different bandwidths and angles are regarded as the feedback to the photonic ISAR imaging system. Furthermore, according to this feedback, we can guide the CNN to converge and recognize based on physical laws, i.e., targets' peculiar features and achieve highly accurate and reliable ATR.

In the proceeding sections, we experimentally demonstrate how to explain the CNN through exploiting the relationship between the photonic ISAR imaging system and the network behaviors to further achieve high-performance ATR. The radio frequency (RF) pulse signals in Fig. 1b are linear frequency modulated (LFM) signals in the frequency intervals of 36-40 GHz and 32–40 GHz with the bandwidths of 4 GHz and 8 GHz, respectively. The pulse width and pulse repetition period of the RF pulse signal are 4 μs and 5 μs, respectively. The detected targets are a fixed-wing aircraft model (Plane), a four-axis drone model (UAV), and a 1:144 Y-20 model (Y20). We use ISAR images collected under two bandwidths of 4 GHz and 8GHz and under random imaging angles to respectively train the CNN, obtaining two trained CNN, named as CNN_4 and CNN_8, respectively. The theories and experiments of the ISAR imaging, the settings of the structure, hyperparameters, and training of theCNN, and the principles of the LRP are illustrated in Methods.

**Results**

We first analyze CNNs' behaviors under different imaging bandwidths. From the ISAR images displayed in Figs. 2 and 3, we can see that ISAR images under 8 GHz are more detailed and more clearly depict the targets' contours and structural features than that under 4 GHz. We use the LRP to explain the reasoning process of the two trained CNNs. Under the conditions of correct identification, the results of the LRP are shown in Fig. 2 by heatmaps, which gives the relevance of the network input and output. The highlighted pixels of the LRP colored by yellow and green or by deep blue positively or negatively contribute to the recognition result of ISAR images by the CNN. Based on the above illustration of the heatmaps, we can observe that the trained CNN_8 rightly recognizes ISAR images of three targets under 8 GHz according to their particular contours and structural features. Concretely, for UAV, the characteristic is several scattering points in the position of wings of UAV and/or in the position of the center point of UAV; for Plane, the characteristic is the profile similar to a triangle; for Y20, the characteristic is the shape looking like a bird. Nevertheless, the features (or pixels) used to rightly recognize ISAR images under 4 GHz by the CNN_4 are blurrier. In fact, we also use other two post-hoc interpretability technologies, that is Integrated Gradient[36] (IG) and Deep Taylor Decomposition[37] (DTD), to confirm the credibility of the interpretability results by the LRP. The interpretability results of the IG and DTD are almost consistent with that of the LRP, which can be found in Supplementary Section.

When the imaging bandwidth decreases from 8 GHz to 4 GHz, the recognition accuracy reduces from 96.00% to 87.78%. We observe the ISAR images, the corresponding network outcomes, and the results of the LRPs, when the two CNNs wrongly recognize targets. Under 8 GHz, the majority of wrong identifications are happened to ISAR images of UAV. Under 4 GHz, the ISAR images of three targets are incorrectly recognized as each other. The results are shown in Fig. 3. The CNN_8 recognizes UAV's ISAR image (Fig. 3a) as Plane's one at the probability of 97.77%. From the result of the LRP of this ISAR image, we can clearly see that the contour of the highlighted pixels colored by yellow and green looks like Plane in Fig. 2. We drop the outcome of the LRP into the CNN_8. The network identifies it as Plane at the probability of 99.41%. This also proves the LRP correctly reflects the critical pixels according to which the CNN_8 gives the recognition result. For the misidentification under 4 GHz, the ISAR image of Y20 in Fig. 3b is wrongly recognized as UAV by the CNN_4 based on the center scattering point. In Fig. 3c, the ISAR image of Plane is incorrectly regarded as that of Y20 and the LRP is non-semantic. We drop the outcomes of the LRP into the CNN_4. The high probability of 99.99% means that the CNN_4 recognizes the ISAR image as Y20 based on unknown laws. In Fig. 3d, the profile of the LRP depicted by yellow and green pixels is faintly similar to a triangle. Hence, under 4 GHz, the low-resolution ISAR images make the NN more difficult to converge according to physical laws such as contour features. As a result, the recognition accuracy is lower and the reasoning process of network is more difficult to be understood. Considering Fig. 2 and Fig. 3, under the same bandwidth, ISAR images of the same target are rightly and wrongly identified. There exist other factors which impact the recognition results.

Next, we explore the CNN_8's behaviors under different imaging angles. We extra collect 414 ISAR images of UAV under 8 GHz. Different imaging angles mean UAV is imaged under different postures. To obtain the imaging angles of UAV, we calculate the root mean square error (RMSE) and the cross-correlation coefficient (the xcorr function in MATLAB 2020a) between two high-resolution range profiles (HRRPs). The one is obtained when UAV is motionless. We experimentally collect 288 samples of HRRP ranging from 0 to 360° at 1.25° interval. The other one is obtained from the first echo of the raw data of the ISAR image when UAV is in motion. Because the number and location of peaks reflect the scattering properties of targets under different postures, the imaging angle of the ISAR image is coarsely mapped to the angle under which the RMSE between two HRRPs becomes smallest. In addition, we use a cross-correlation coefficient to measure the similarity between the above two HRRPs. Figure 4 shows six ISAR images under different imaging angles. The blue curves of each subfigure are the HRRPs of UAV at the top left corner of each subfigure. The red curves are the HRRPs obtained from the raw data of six ISAR images, which have the smallest RMSE with the corresponding blue curves. (Some of them are not similar. The reasons are presented in Discussion.) Thus, we mark the 185.00°, 46.25°, 97.50°, 148.75°, 172.50°, and 112.50° as the imaging angles of the six ISAR images. The real gestures of UAV are generally inconsistent with the ISAR images, which is reasonable. Because the ISAR image reflects the projection of the target under Plane constituted by radar line of sights (RLOS) and its perpendicular direction (See the ISAR imaging principles in Method). From the perspective of the ISAR images, Figures 4a and 4b intuitively look like UAV and their LRPs show a few of scattering points in the position of wings of UAV and/or the center point of UAV. Figures 4c and 4d resemble Plane and Y20 and their LRPs display the profile similar to a triangle and the shape looking like a bird, respectively. The symbolic triangle and bird are critical features based on which the CNN_8 recognizes ISAR images as Plane and Y20, respectively. Figures 4e and 4f correspond to the cases under certain imaging angles, which are hardly recognizable. The LRP in Fig. 4e shows that the CNN_8 powerfully covers the latent and critical features from the ISAR image and thus correctly identifies it. But, in the LRP of Fig. 4f, the contour of the highlighted pixels colored by yellow and green seems like Plane in Fig. 2 and the corresponding ISAR image is wrongly recognized. Hence, the angles under which UAV are imaged influence on the characteristics which the ISAR images embody. The embodied features impact on the accuracy and reliability of network recognition results.

Furthermore, we analyze the relationship between recognition results and imaging angles. The recognition accuracy of the 414 ISAR images of UAV is 85.27%. The 414 ISAR images altogether cover 161 imaging angles within the 288 angles from 0 to 360° at 1.25° interval. Table 1 presents the statistical results of network recognition under different imaging angle ranges. In Table 1, we mark angle ranges with bold font under which the difference among the imaging angles of each ISAR image do not surpass 1.25° and the network recognition accuracies are larger than 85.27%. Moreover, we mark angle ranges with black underlines, each angle under which there is more than one ISAR image. Thus, the recognition results of ISAR images under angle ranges with bold font and black underlines are robust. Under angle ranges of 27.50°-33.75°, 165.00°-178.75°, and 261.25°-270.00°, we can achieve high-accuracy ATR with the accuracies of 95.45%, 90.48%, and 94.44%, respectively. By observing the ISAR images and their LRPs under these angle ranges, we have found that the majority of ISAR images explicitly or implicitly embody peculiar features of UAV and the CNN_8 depends on the several scattering points in the position of wings of UAV and/or the center point of UAV to correctly identify these ISAR images, which can be found in Supplementary Section. We regard the three angle ranges as appropriate imaging angle ranges, because under these angle ranges the process of the ISAR imaging keeps UAV's unique features and thus network recognition results are more high-resolution, explainable, and reliable. Note that there also exist other appropriate imaging angles outside the three angle ranges with bold font and black underlines. The ISAR images under these imaging angles also embody peculiar features of UAV and the network rightly recognizes them. In view of these angles being discrete, we do not give specific values of angles here.

Finally, we intuitively observe network behaviors to ISAR images obtained under different imaging bandwidths and imaging angles with the assistance of the t-SNE. The ISAR images used in Figs. 5a and 5b are collected at the same day with the training dataset which is used to train the CNN_4 and CNN_8. Moreover, to only consider the influence of imaging bandwidths, ISAR images in Figs. 5a and 5b are collected under random imaging angles. In addition, all ISAR images in Fig. 5 do not include the ISAR images in the training datasets. Comparing Fig 5b with Fig. 5a, the results of the t-SNE in the network output layer on different targets' ISAR images collected under 8 GHz are separated further than that under 4 GHz. Moreover, the results of the t-SNE in the network output layer on the same target's ISAR images collected under 8 GHz are gathered more compactly than that under 4 GHz. Figure 5c shows the t-SNE of the additional 414 ISAR images of UAV which are collected at another day with the training datasets. The purple, cyan, and yellow stars in Fig. 5c present the results of t-SNE in the network output layer on ISAR images under angle ranges of 27.50°-33.75°, 165.00°-178.75°, and 261.25°-270.00°, respectively. We can see that for most ISAR images of UAV under the three imaging angle ranges, the features in the network output layer after dimensionality reduction by the t-SNE are far from that of Y20 and Plane.

All in all, the ISAR images obtained under small bandwidths cannot clearly depict the targets' features and the degree of discrepancy among targets decreases, which makes it harder for the CNN to converge and recognize according to physical laws, i.e., targets' peculiar features. As a result, the recognition accuracy is lower and the reasoning process of network is more difficult to understand. That is the reliability is lower. When targets are imaged under the broader bandwidths and appropriate imaging angles, the physical process of ISAR imaging keeps targets' peculiar features and the discrepancy among targets is larger. The NN recognizes ISAR images depended on targets' features. The reasoning process is more semantic and reasonable. Therefore, improving the imaging bandwidth and imaging under appropriate angles bring to the high-accuracy and reliable ATR based on the photonic ISAR imaging system.

**Discussion and Conclusions**

The imaging angles are marked a little imprecisely due to the limited experiment conditions. We manually rotate the rotation table at 1.25° interval in a counterclockwise direction, which almost approaches the maximum limitation of precision. During this process, angle errors among those 288 HRRPs may exist due to the visual errors. The ISAR images are obtained when the rotation table quickly automatically rotates counterclockwise. The HRRPs of the ISAR images may not be within the 288 marked angles. Hence, sometimes, the similarities measured by the cross-correlation coefficient between the HRRP of ISAR images and the HRRP mapped according to the smallest RMSE values are not very high. The additional reason causing this problem may be a little difference of system states due to those data being collected at different days. The difference influences the amplitude of each peak of the HRRPs but hardly impacting on the angle marking results. Because angle marking is based on the positions of each peak. In reality application, we can obtain target's HRRPs at smaller angle intervals with the help of angle measurement devices such as gyroscopes and achieve more accurate angle marking of the ISAR images.

For the ATR based on ISAR imaging, when a target is imaged under certain inappropriate angles, its ISAR images look like other targets' or include the latent features of other targets. The network wrongly recognizes them with high probability. Moreover, the corresponding LRPs show obvious features of other targets. Under this circumstance, we will be confused by the network output, because we do not know the results is right or wrong. This is a huge challenge which exists in all ATR tasks. We can collect several ISAR images of the same target under different imaging angles in a short time and combine all the recognition results of these ISAR images together to get the final results. Imaging target under appropriate angle from the results of angle marking is feasible to reduce the risk of wrong identification. Combined the different target themselves physical characteristics such as electromagnetic characteristics and deep comprehension of the ISAR imaging principles might bring a potential solution. Some methods from the software level like Ref.[47] can also be explored to resolve this issue.

Based on the thinking of our work and the early knowledge accumulation of NN, we think that for the intelligent recognition by NN, besides the ATR based on ISAR images, the network acquires several particular features of each category during the training phase; these features contribute to the recognition of category with different weights; in testing phase, the network recognizes images as a certain category according to the features exposed or hidden in the images; which features the image embodies largely depends on the physical process of imaging; when the images of one category, being collected under several conditions, include the critical features which are largely contributed to the recognition of another category, the network will wrongly identify it. Thus, exploring how the physical process of the imaging impacts on the features of images and further influences the network behaviors, is significant for high-performance ATR.

In summary, to achieve high-accuracy and reliable ATR based on ISAR images, we investigate the relationship between the photonic ISAR imaging system and the network behaviors from two aspects, that is imaging bandwidths and imaging angles. Combined the results of network recognition and the corresponding LRPs with the visualization of the t-SNE in the network output layer, we have found that when the imaging bandwidth is broader, the ISAR images embody more detailed information and the discrepancy of ISAR images among different targets are more obvious, which favors the network converging according to physical laws and recognizing ISAR images rightly based on targets' peculiar features. Moreover, when the targets are imaged under appropriate angles, the ISAR images keep targets' unique characteristics and there exists large discrepancy among ISAR images of different targets. Therefore, the recognition results of ATR based on the photonic ISAR imaging system can be high-resolution and reliable by increasing imaging bandwidth and imaging under appropriate angles. This is the first time to explaining the NN for ATR by exploring the imaging physical process on network behaviors according to the domain knowledges of imaging and further polish the performance of network output. This may bring new insights for other researchers related in ATR to achieve high-performance recognition and in explainable AI to explain and explore the principles of AI for the accomplishment of better intelligent abilities. In terms of the future directions, we will further deeply explain AI from the insight of differential.

**Materials and Methods**

**ISAR imaging principles and experiments**

The RF pulse signal transmitted by the radar's transmitting horn antenna reflects on the surface of targets, forming echoes. The echoes received by the radar's receiving horn antenna. There are several scattering points distributed in different locations of the target's surface. Hence, the echo of each scattering point with different distances to radar are received at different times. We operate pulse compression on those echoes, obtaining the HRRP in range direction. The location of each peak in the HRRP reflects the distance from scattering points to radar. The amplitude of each peak means the reflective intensity of scattering points at the same distance, to some extent. Consequently, the HRRP reflects the projection of targets in range direction, that is RLOS. The range resolution $\delta_r$ depends on the bandwidth $B$ of the RF pulse signal, which can be expressed mathematically as:

$$\delta_r = \frac{c}{2B} \tag{1}$$

where $c$ means the speed of light. To obtain the two-dimensional image, we also need take the projection of targets in the azimuth direction, that is the perpendicular direction of the RLOS, into account. We consider the range-azimuth rotation model of ISAR imaging. When the target performs uniform circular motion, scattering points distributed in the azimuth direction and with the same distance in range direction from radar, have different velocities in the direction of the RLOS, causing the different doppler frequency shifts. The doppler frequency shift $f_d$ reflects the distance from scattering points to the center of the target in azimuth direction $y$, which can be expressed as:

$$f_d = \frac{2w}{\lambda} y \tag{2}$$

where the $w$, and $\lambda$ mean the circumferential velocity and the wavelength of the RF signal, respectively. In the center point of target, the $f_d$ is zero. When the target rotates several angles, the radar transmits several pulses of RF signal and receives several pulses of echoes. Based on the phase accumulation, we can obtain the HRRP in azimuth direction. The

azimuth resolution $\delta_r$ can be expressed mathematically as:

$$\delta_z = \frac{c}{2\theta f_c} \tag{3}$$

where $f_c$ and $\theta$ mean the center frequency of the RF signal and the total rotation angles of the rotating table during the pulse accumulation time. More detailed ISAR imaging theory can refer to Ref.[48,49]. There are several algorithms[50,51] can be adopted to process pulses echoes and generate the ISAR images.

In our experiments, the PRS in Fig. 1b is a photonic analog-to-digital converter (PADC) with the sampling rate of 40 GSa/s. The PADC directly receives the RF pulse signal with the high frequency and big bandwidth without complex frequency mixing. The concrete experimental setup of PADC is displayed in Fig. S1 (see Supplementary Section). In the experiment with the bandwidth of 8 GHz, an arbitrary waveform generator (AWG, KEYSIGHT M8195A) generates a baseband LFM pulse signal with frequency from 1 GHz to 9 GHz. A microwave signal generator (MSG, KEYSIGHT N5166B) generates a carrier signal whose frequency is 31 GHz. We mix the LFM pulse with the carrier signal via an electrical mixer (MIXER, Marki MM1-1140H), obtaining a RF LFM pulse signal with the frequency from 32 GHz to 40 GHz. In the experiment with the bandwidth of 4 GHz, the frequency of the baseband LFM pulse signal is from 2 GHz to 6 GHz and the frequency of the carrier signal is 34 GHz. We mix the two signals and obtain a RF pulse signal with the frequency from 36 GHz to 40 GHz. The pulse width of the RF signal is 4 µs and the repetition frequency is 5 µs. The RF pulse signal is amplified by an amplifier (AMP1, CONNPHY CMP-18G-40G-3021-K) and then transmitted by a horn antenna (Tx, INFOMW LB-28-25-C2-KF). The RF pulse signal reflects on the moving target which is fixed on the rotation table. The rotation table rotates counterclockwise. The rotating speed is around 2r/s. The reflected LFM pulses echoes are received by another horn antenna (Rx, INFOMW LB-28-25-C2-KF) and is amplified by another amplifier (AMP2, CONNPHY CLN-36G-40G-3030-K). The echoes received by the Rx are modulated into the pulse light sequence through a Mach-Zehnder modulator (MZM, EOSPACE AX-0MVS-40-PFA-PFA-LV). The pulse light sequence is emitted by an actively mode-locked laser (AMLL, PSL-40-TT) and its repetition frequency is set to be 40 GHz. The modulated pulse light is separated into two-way light pulse sequences with a sampling rate of 20 GSa/s after a dual-output MZM (DOMZM, EOSPACE, AX-1× 2-0MxSS-20-SFU-LV). Two photodetectors (PD EOT ET-3500F) array with a bandwidth of ~15 GHz are used to convert the pulse light into the electrical sequences. The baseband signal to which the RF signal is down converted by mixing with the pulse light, are digitized by the Oscilloscope (OSA, KEYSIGHT DSO-S 804A). The sampling rate and bandwidth of the OSC are of 20 GSa/s and 10 GHz, respectively. The time storage length of the OSA is set to 20 ms. In digital signal processing (DSP), we use the back projection (BP) algorithm[50,51] to obtain the ISAR images.

We altogether perform three experiments. The first experiment obtains 300 ISAR images of the three targets, i.e., Y20, Plane, and UAV, under random imaging angles and imaging bandwidth of 8 GHz, each target 100. The second experiment obtains 240 ISAR images of the three targets under random imaging angles and imaging bandwidth of 4 GHz, each target 80. The third experiment obtains 414 ISAR images of UAV under 161 imaging angles and imaging bandwidth of 8 GHz.

**The structure, parameters, and training of the CNN**

The CNN in this study includes 5 convolutional (Conv.) layer, 5 maximum pooling layers, 3 fully connected (FC) layers. The sizes of the kernels in Conv. layers and maximum pooling layers are 3*3 and 2*2, respectively. The numbers of the feature map in the Conv. layers are 32,64,64,128, and 32. The numbers of neurons in the FC layers are 300, 100, and 3.

The activation function is Relu. In the second FC layer, we adopt L1 regularization with the parameter of 0.001. The loss function is categorical crossentropy. The batch size and total training epoch are 32 and 200, respectively. The numbers of training datasets under both 4 GHz and 8 GHz are 150, each target 50. The numbers of testing datasets under 4 GHz and 8 GHz are 90 and 150, each target 30 and 50, respectively. The recognition accuracy of the CNN_4 and the CNN_8 is 87.78% and 96.00%, respectively. The losses and accuracies curves of the CNN_4 and the CNN_8 are displayed in Fig. S4 (see Supplementary Section). The CNN model is coded in Python 3.6 and is implemented in TensorFlow 1.15.

**Principles of the LRP**

The LRP computes the relevance scores of each neuron. In the output layer, the relevance scores of every neuron are typically taken as the actually pre-activation value corresponding to the class which are obtained from the forward inference of an input through the trained CNN. Relevance scores are propagated to the preceding layers and the sum of the relevance scores in each layer is constant, which can be expressed as:

$$f(x) = \cdots = \sum_{i \in l+1} R_i^{(l+1)} = \sum_{i \in l} R_i^{(l)} = \cdots = \sum_i R_i^{(1)} \qquad (4)$$

where $f(x)$ is the real-valued prediction output, and $R_i^l$ is the relevance score for neuron $i$ in the layer $l$. For fully-connected and convolutional layers, there are several ways to propagate relevance to the previous layer while satisfying equation (4)[35]. In our work, we adopt the z-rule[53], for which the propagation of relevance is computed as:

$$R_i^l = \sum_j \left( \frac{z_{ij}}{\sum_{i'} z_{i'j} + \varepsilon \cdot sign(\sum_{i'} z_{i'j})} \right) R_j^{l+1} \qquad (5)$$

where $sign$ is the SIGN function, $\varepsilon$ is for the numerical stability, and $z_{ij}$ is the neuron output before non-linear activation. $z_{ij} = a_i^{(l)} w_{ij}^{(l,l+1)}$, where $a_i^{(l)}$ is the activation of neuron $i$ in the layer $l$ and $w_{ij}^{(l,l+1)}$ is the weight connecting neurons $i$ and $j$.

Nowadays, there are several soft packages which include numerous post-hoc interpretability methods. We can directly use the LRP by call the related library function. In our work, we use the presented library iNNvestigate[54] which can be simply installed as Python package and contains documentation for code and applications. In addition, iNNvestigate is available at repository: https://github.com/albermax/innvestigate.

**Acknowledgements**
This work is supported in part by National Key Research and Development Program of China (Program No. 2019YFB2203700) and National Natural Science Foundation of China (Grant No. T2225023).

**Author contributions**
W. Z. and X. Z. proposed the novel scheme of explaining the CNN for ATR. A. D. designed the ISAR imaging experimental setup. X. Z., A. D., L. Z., and S. H contributed to the ISAR imaging experiments. X. Z. and A. D. processed ISAR imaging data. X. Z. trained the CNN and operated the t-SNE. X. Z. and Y. H. performed the LRP. X. Z. analyzed the results of network output, LRP, and t-SNE. X. Z., S. X., and W. Z. prepared the manuscript. W. Z. initiated and supervised the study.

**Data availability**

All data in the main text or the supplementary materials are available from the corresponding author upon request.

**Code availability**

All the codes used for this study are available from the corresponding author upon request.

**Competing interests**

The authors declare no competing interests.

**Supplementary information** is available for this paper at https://doi.org/10.1038/xxxxxxxx.

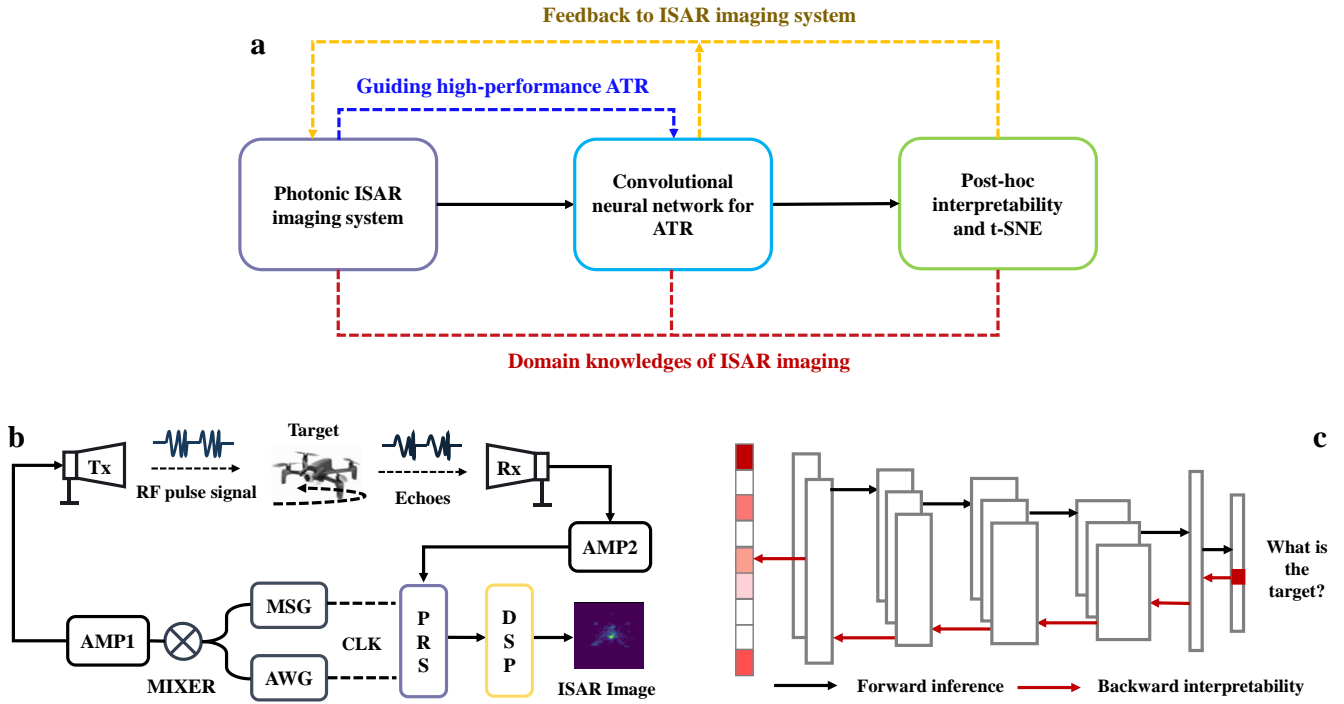

**Fig. 1 Schematic of the achievement of the high-performance ATR based on the photonic ISAR imaging system via explaining the NN. a**, Basic principles of how to explain the intelligent behaviors of the CNN and guide a high-resolution and reliable ATR. First, the photonic imaging system produces ISAR images. A portion of images are used to train the CNN. Next, a post-hoc interpretability technology is adopted to obtain the relevance between each pixel of ISAR image and its network recognition result and a dimensionality reduction algorithm is utilized to visualize the features in the network output layer. Then, we change the imaging process based on domain knowledges of the ISAR imaging to obtain different ISAR images. The network outputs of these images, the corresponding outcomes of the LRP, and the visualization of the t-SNE are analyzed, forming the feedback to the photonic ISAR imaging system and guiding us to obtain high-performance ATR which recognition results are high-accuracy and reliable. **b**, Simplified architecture of the photonic ISAR imaging system. The detailed experimental setup of this system can be found in Supplementary Section. The RF pulse signal is transmitted by the antenna (Tx) and reflects on the surface of the target. The radar echoes are received by the photonic receiving system (PRS). After the digital signal processing (DSP), we obtain ISAR images. **c**, Illustration of the post-hoc interpretability technologies. In the forward inference phase, the trained CNN gives the recognition outcome. In the backward phase, the LRP gives the relevance between the input and network recognition outcome. t-SNE: t-distribution stochastic neighbor embedding, a dimensionality reduction algorithm.

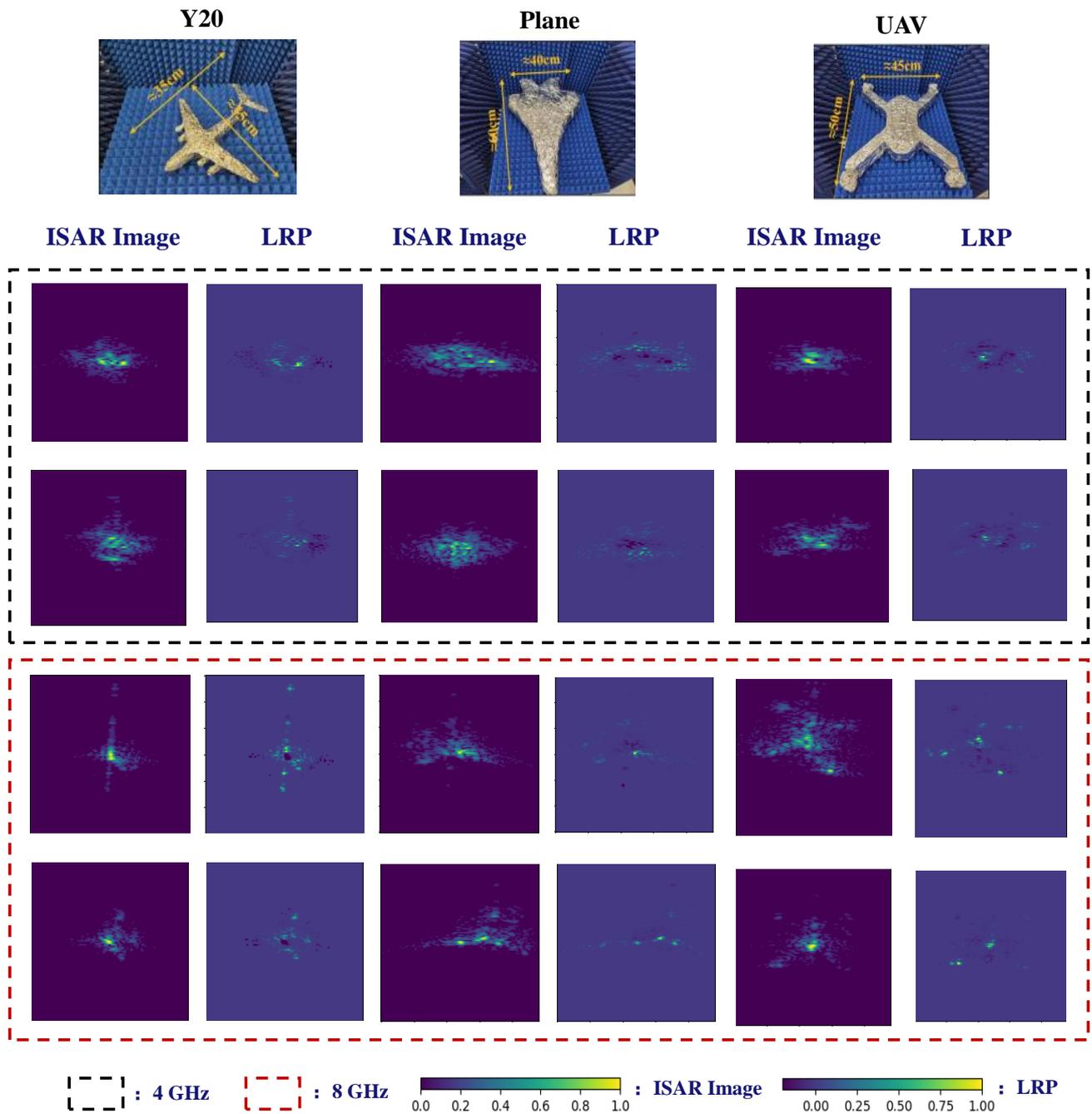

**Fig. 2 ISAR images which are correctly recognized and their corresponding LRPs under different bandwidths.** The ISAR images are normalized to the range of 0-to-1. The results of the LRP show the relevance between each pixel of the images and their corresponding recognition results, which are displayed by heatmaps whose color scale is from -0.20 to 1.00. The positive and negative values of the result of the LRP which are depicted by colors closing to yellow and blue, represent the pixels of an image support and oppose the trained CNN to recognize it as the network output, respectively.

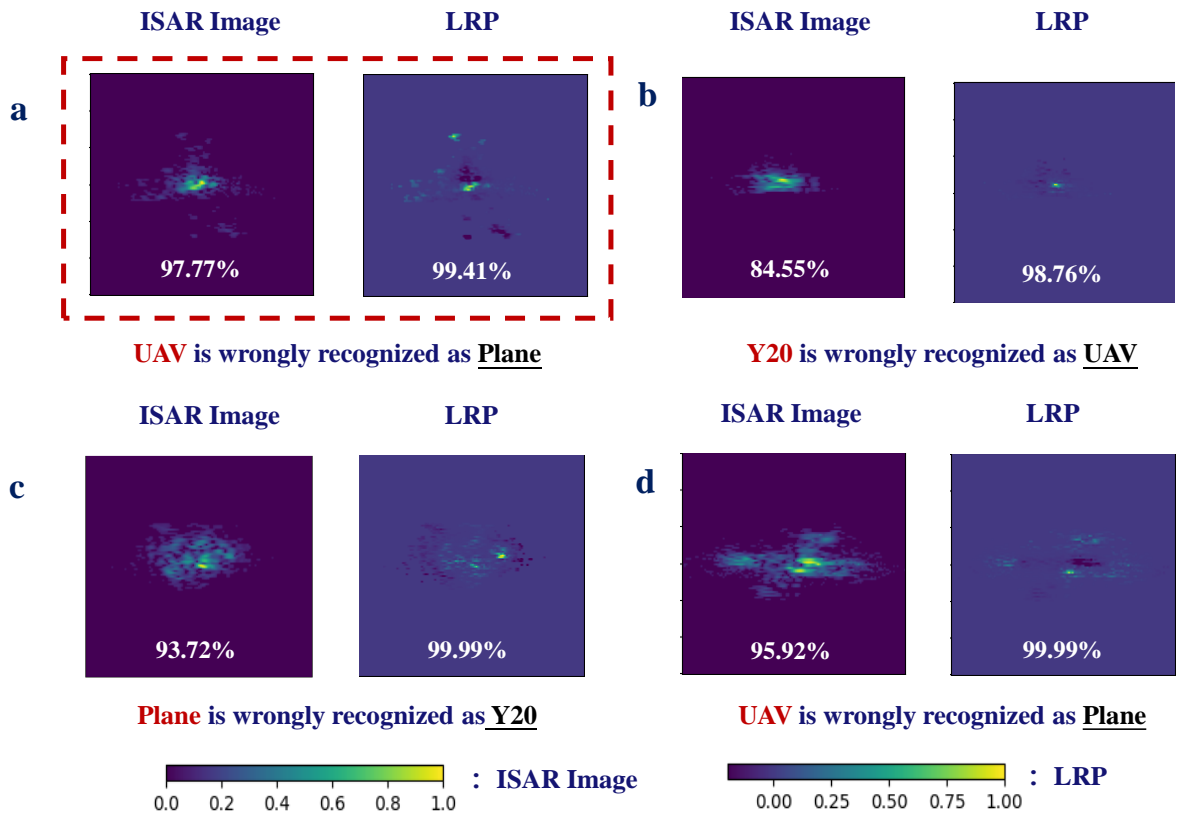

**Fig. 3 ISAR images which are wrongly recognized and their corresponding LRPs under different bandwidths.** Four examples of misidentification are shown. Expect that the one in the red dotted box is under 8 GHz, the other three are under 4 GHz. The outcomes of the LRP are also dropped into the CNN. The white numbers show the probabilities that the CNN recognizes them as categories highlighted by black underlines.

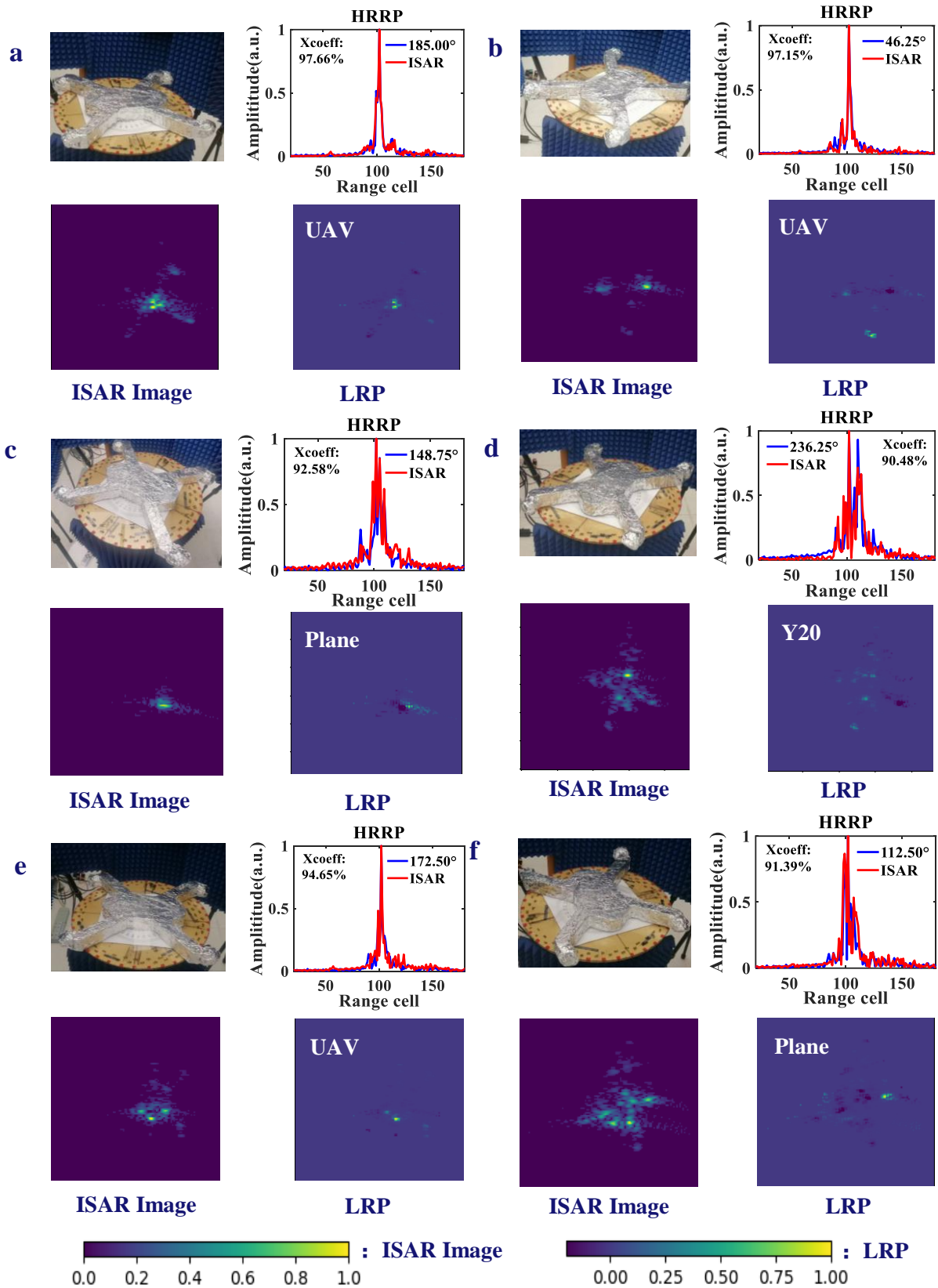

**Fig. 4 ISAR images of UAV and their corresponding LRPs under the bandwidth of 8 GHz and different imaging angles.**
**a**, **b**, The ISAR images which look like UAV are recognized rightly. **c**, **d**, Two ISAR images which look like Plane and Y20 are recognized wrongly. **e**, **f**, Two hardly recognizable ISAR images. **f,** The ISAR image is wrongly recognized as Plane. HRRP: the high-resolution range profile. The "Xcoeff" means the coefficient of covariance, which presents the similarity between the two HRRPs.

Table 1: Recognition accuracies under different angle ranges

| Angle Range | Error/Total | Acc. (%) | Angle Range | Error/Total | Acc. (%) | Angle Range | Error/Total | Acc. (%) |
|---|---|---|---|---|---|---|---|---|
| [0.00°, 10.00°] | 3/17 | 82.35 | [118.75°, 123.75°] | 2/14 | 85.71 | [236.25°, 238.75°] | 2/4 | 50.00 |
| **[13.75°, 22.50°]** | **0/8** | **100.00** | [126.25°, 128.75°] | 1/8 | 87.50 | [241.25°, 250.00°] | 3/8 | 62.50 |
| **[27. 50°, 33.75°]** | **2/44** | **95.45** | [135.00°-140.00°] | 0/5 | 100.00 | [251.25°, 258.75°] | 0/5 | 100.00 |
| [36.25°, 42.50°] | 4/17 | 76.47 | [145.00°-148.75°] | 2/6 | 66.67 | **[261.25°, 270.00°]** | **1/18** | **94.44** |
| [45.00°, 48.75°] | 1/8 | 87.50 | [153.75°, 158.75°] | 0/4 | 100.00 | [272.5°, 280.00°] | 5/26 | 80.77 |
| [51.25°, 58.76°] | 3/12 | 75.00 | **[165.00°, 178.75°]** | **4/42** | **90.48** | [291.25°, 298.75°] | 1/11 | 90.91 |
| [62.50°, 65.00°] | 0/6 | 100.00 | [181.25°, 188.75°] | 0/15 | 100.00 | [301.25°, 303.75°] | 1/6 | 83.33 |
| [73.75°, 76.25°] | 1/11 | 90.90 | [191.25°, 200.00°] | 1/6 | 83.33 | [312.50, 316.25°] | 0/6 | 100.00 |
| [78.75°, 88.75°] | 4/9 | 55.56 | [201.25°, 205.00°] | 4/7 | 42.86 | [350.00°, 355.00°] | 0/12 | 100.00 |
| [97.50°, 105.00°] | 3/19 | 84.21 | [211.25°, 215.00°] | 1/3 | 66.67 | | | |
| **[111.25°, 116.25°]** | **1/12** | **91.67** | [217.50°, 222.50°] | 3/21 | 85.71 | | | |

Acc. is the abbreviation of accuracy. The total accuracy of the 414 ISAR images is 85.27%. The angle range marked with bold font means that during this angle range, the angle difference between each sample do not surpass 1.25° and the accuracies are larger than 85.27%. In the angle ranges with black underlines, almost all angles have more than one example.

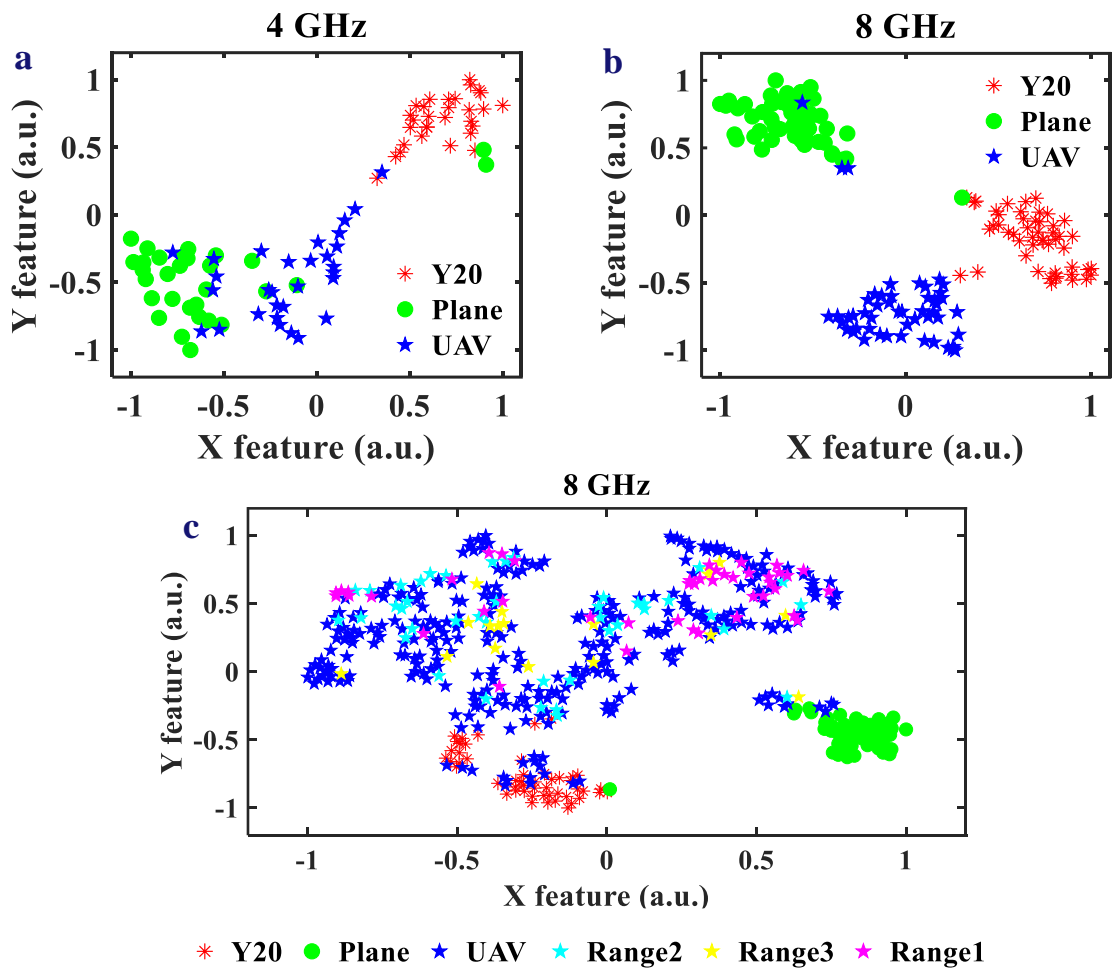

**Fig. 5 Results of the t-SNE of the network output layer of ISAR images under different imaging bandwidths and imaging angles. a**, **b**, Results of ISAR images under 4 GHz and 8 GHz, respectively. **c**, Results of additional 414 ISAR images. Ranges 1-3 present the three imaging angle ranges of 27.50°-33.75°, 165.00°-178.75°, and 261.25°-270.00°, respectively.

# Supplementary information

## S1. Experimental setup of photonic ISAR imaging system

Figure S1 displays the experimental setup of a photonic ISAR imaging system. The receiver is a photonic analog-to-digital converter (PADC) system which can receive high frequency and big bandwidth RF signals without complex operation of frequency mixing. In our experiment, the antennas (Tx and Rx) look down the target. The target is fixed on a rotation table which rotates in the counterclockwise direction.

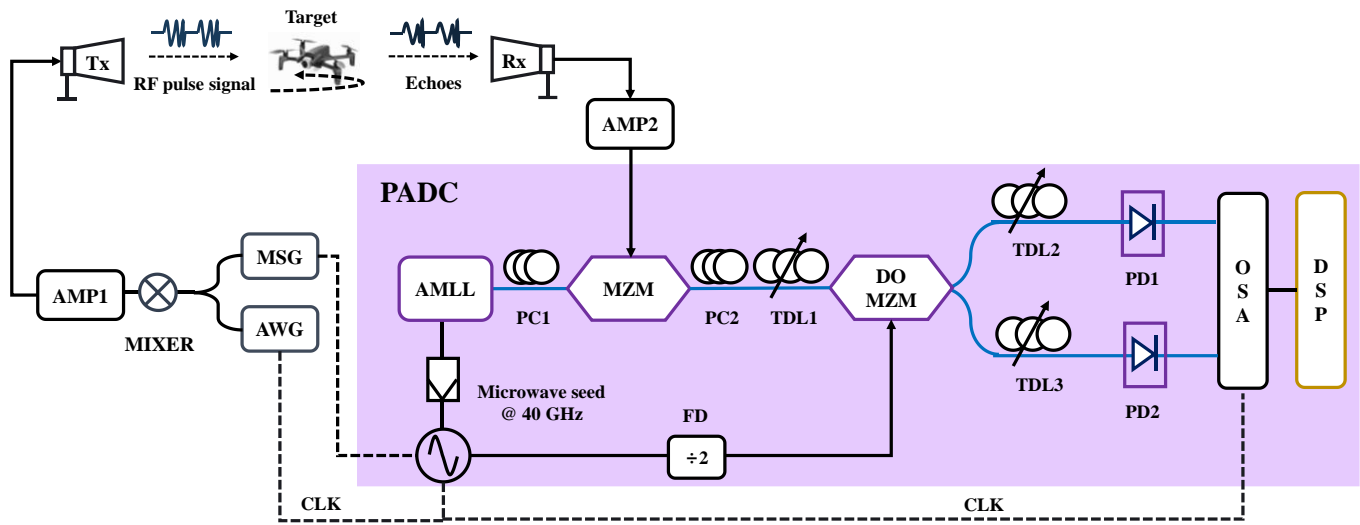

**Fig. S1 The experimental setup of photonic ISAR imaging system.** The architecture of PADC is highlighted with purple fill box. Tx: transmitting horn antenna; Rx: receiving horn antenna. AWG: arbitrary waveform generator; MSG: microwave signal generator; AMP: amplifier; AMLL: actively mode-locked laser; PC: polarization controller; MZM: Mach-Zehnder modulator; TDL: tunable delay line; DOMZM: dual-output Mach-Zehnder modulator; PD: photodetectors; FD: frequency divider; OSA: oscilloscope; DSP: digital signal processing; CLK: the reference clock.

## S2. The result of three post-hoc interpretability technologies

We use three interpretability technologies to exploit how the CNN_8 recognizes different targets. In Fig. S2, the results of the DTD, IG, and LRP reflect the relevance between pixels of ISAR images and the outcome of network recognition. The larger the relevance is, the brighter the corresponding pixel is highlighted. Comparing the results of the DTD with their corresponding ISAR images, we can see that the CNN neglects many unimportant pixels which depict the surfaces of targets and depends on many pixels which depict the contours of targets to recognize different targets. The values of the results of the DTD are always positive due to the property of the DTD's principles which is illustrated in Ref.[37]. Thus, the results of the DTD reflect the influence of input features. Whereas we do not know it is positive or negative. The results of the IG and LRP are almost same. Among these results of interpretability, we roughly sketch the contours of the pixels highlighted by yellow color via the white dotted lines. The white contours largely outline the shape and structure of targets. This indicates that the CNN_8 distinguishes different targets according to their shapes and structures and confirms the credibility of the interpretability results by the LRP.

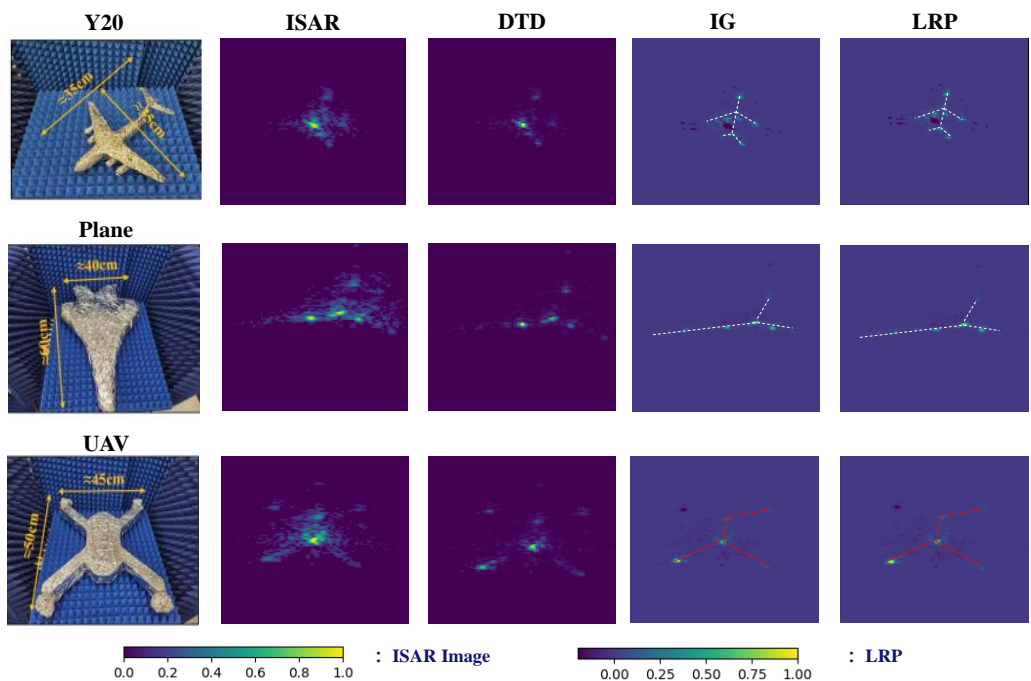

**Fig. S2 Results of three post-hoc interpretability technologies.** IG: Integrated Gradient[36]; DTD: Deep Taylor Decomposition[37] (DTD); LRP: layer-wise relevance propagation (LRP)[35].

**S3. ISAR images and their corresponding LRPs under the three appropriate imaging angle ranges**

Figure S3 displays 18 examples of ISAR images and their corresponding LRPs under the three appropriate imaging angle ranges of 27.50°-33.75°, 165.00°-178.75°, and 261.25°-270.00°, each angle range 6 examples. The posture of UAV under 0° is shown in the top of Fig. S3. The UAV is fixed on the rotate table which rotates in the counterclockwise direction. For each ISAR image, we mark its imaging angle and cross-correlation coefficient in Fig. S3. Form Fig. S3, we can see that the majority of ISAR images explicitly or implicitly embody peculiar features of UAV and the CNN_8 depends on the several scattering points in the position of wings of UAV and/or the center point of UAV to correctly identify these ISAR images. Thus, the recognition results are high-accuracy and reliable.

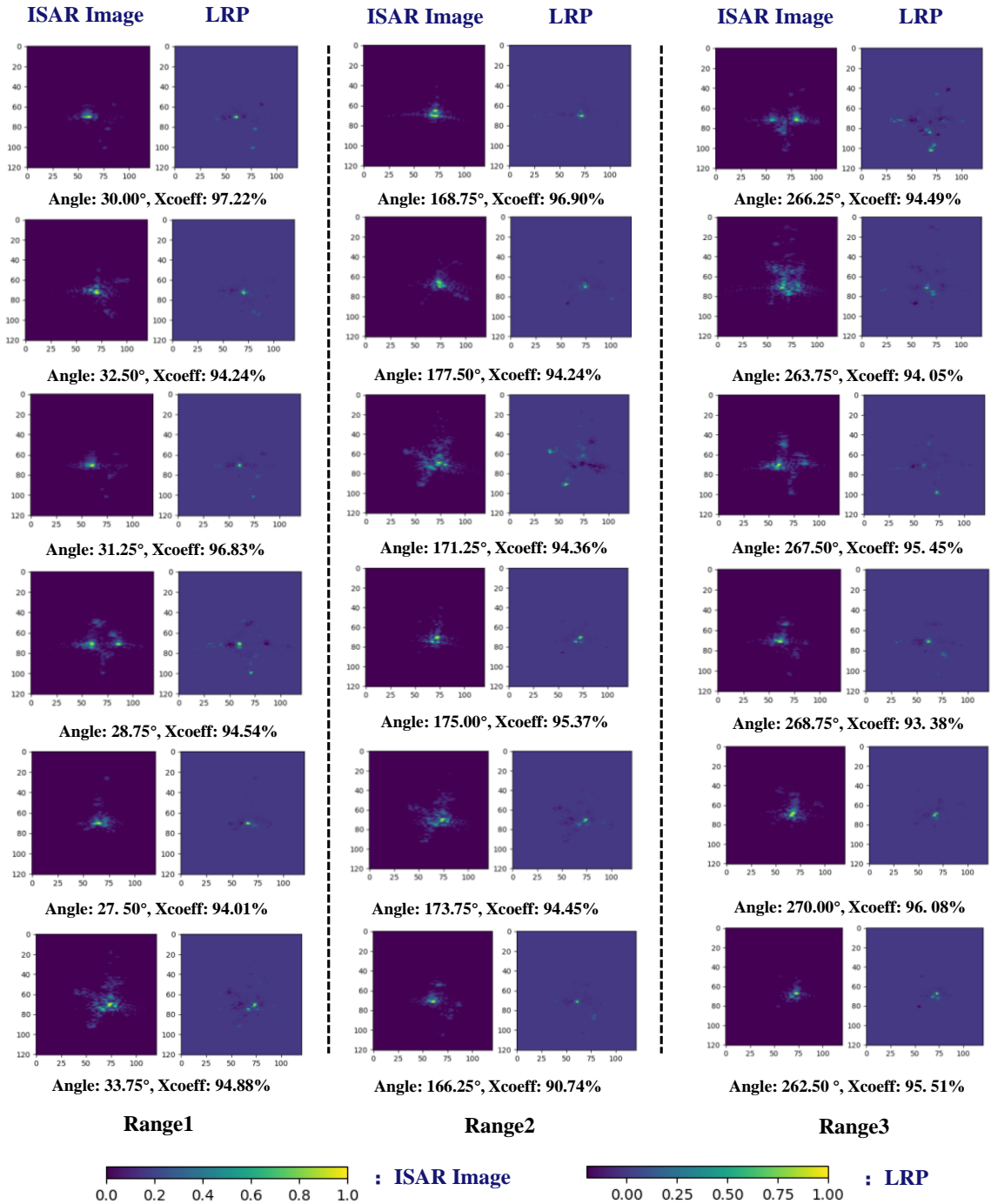

Fig. S3 ISAR images and their corresponding LRPs under the three appropriate imaging angle ranges. Ranges 1-3 present the three imaging angle ranges of 27.50°-33.75°, 165.00°-178.75°, and 261.25°-270.00°, respectively.

## S4. The loss and accuracy curves of the CNNs

We use ISAR images of three targets, each target 50 examples, obtained under the bandwidths of 4 GHz and 8 GHz to train two CNNs. The two CNNs trained with ISAR images under 4 GHz and 8 GHz are named as CNN_4 and CNN_8, respectively. From Fig. S4, we can know that both the CNN_4 and CNN_8 are well converged and without overfitting. The testing losses of the CNN_4 and CNN_8 decrease to 0.76 and 0.17, respectively. The testing accuracies of the CNN_4 and CNN_8 are 87.78% and 96.00%, respectively.

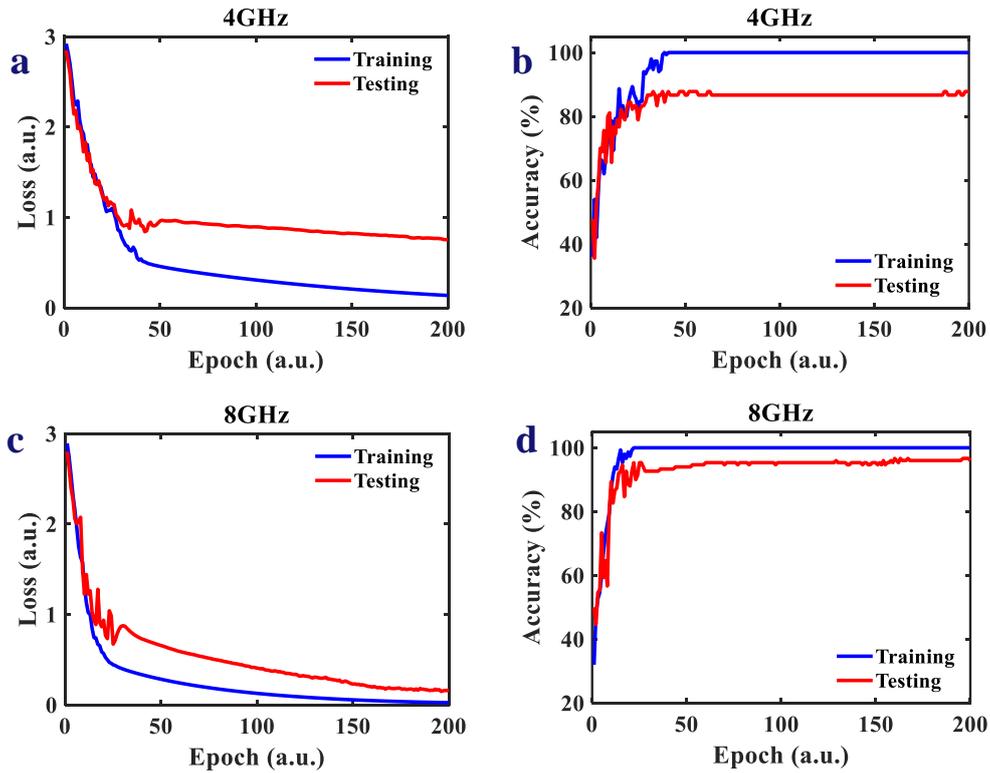

**Fig. S4 The loss and accuracy curves of the CNN_4 and CNN_8. a, b,** The loss and accuracy curves of the CNN_4. **c, d,** The loss and accuracy curves of the CNN_8. The testing loss of the CNN_8 decreases to smaller values than that of the CNN_4 at the same epochs. The testing accuracy of the CNN_8 is higher than that of the CNN_4.